\documentclass{article}

\usepackage{arxiv}
\usepackage{graphicx}
\usepackage[utf8]{inputenc} 
\usepackage[T1]{fontenc}    
\usepackage{hyperref}       
\usepackage{url}            
\usepackage{booktabs}       
\usepackage{amsfonts}       
\usepackage{nicefrac}       
\usepackage{microtype}      
\usepackage{lipsum}		
\usepackage{subfig}
\title{Spin-momentum locked modes on anti-phase boundaries in photonic crystals}

\author{
  Xianghong Kong \\
  Department of Electronic Engineering\\
  Shanghai Jiao Tong University\\
  Shanghai 200240, China \\
   Department of Electrical and Computer Engineering\\
  University of California, San Diego\\
  California 92093, USA\\
  \texttt{klovek@sjtu.edu.cn} \\
   \And
 Yun Zhou \\
  Department of Mechanical and Aerospace Engineering\\
  University of California, San Diego\\
  California 92093, USA \\
  \texttt{yuz421@eng.ucsd.edu} \\
  \AND
  Gaobiao Xiao \\
 Department of Electronic Engineering\\
Shanghai Jiao Tong University\\
Shanghai 200240, China \\
  \texttt{gaobiaoxiao@sjtu.edu.cn} \\
\And
 Daniel F. Sievenpiper \\
 Department of Electrical and Computer Engineering\\
University of California, San Diego\\
California 92093, USA\\
 \texttt{dsievenpiper@eng.ucsd.edu} \\
}

\begin{document}
\maketitle

\begin{abstract}
An anti-phase boundary is formed by shifting a portion of  photonic crystal lattice along the direction of periodicity. A spinning magnetic dipole is applied to excite edge modes on the anti-phase boundary. We show the unidirectional propagation of the edge modes which is also known as spin-momentum locking. Band inversion of the edge modes is discovered when we sweep the geometrical parameters, which leads to a change in the propagation direction. Also, an optimized source is applied to excite the unidirectional edge mode with high directivity.
\end{abstract}

\keywords{Spin-momentum locking \and Anti-phase boundary \and Photonic crystal waveguides}

\section{Introduction}

The quantum spin-Hall effect indicates that the spin of the electron is locked to the direction of propagation \cite{kane2005quantum}. The $Z_2$ index, or the spin Chern number which is a topological invariant of the given quantum system is defined to verify if the spin Hall conductance exists on the edge of the bulk material \cite{kane2005z,sheng2006quantum}. After introducing the $Z_2$ topological index to analyze the system, a variety of unidirectional edge modes in quantum systems were discovered \cite{konig2008quantum,kuemmeth2008coupling,soumyanarayanan2016emergent}. By analogy with the quantum spin-Hall effect of electrons, spin-momentum locking phenomena can also be found in photonic topological insulators \cite{khanikaev2013photonic,wu2015scheme,ma2017scattering,ozawa2019topological,hafezi2011robust}. The direction of propagation is still used to define 'momentum' of the light while the concept of 'spin' is not as clear as the spin of the electron. It may refer to the bonding (antibonding) states of electric and magnetic fields \cite{khanikaev2013photonic}, left-hand (right-hand) circular polarizations of electric fields \cite{wu2015scheme}, and clockwise (anticlockwise) circulations of coupled resonator optical waveguides \cite{hafezi2011robust}. 

Spin-momentum locked edge modes can also be discovered in trivial optical systems without topological properties, such as photonic crystal waveguides \cite{sollner2015deterministic,coles2016chirality,young2015polarization}, surface plasmon polaritons \cite{van2016universal,rodriguez2013near}, and even dielectric waveguides \cite{rodriguez2013near}. A pair of orthogonal dipoles with $\pm\pi/2$ phase differences which represent opposite spin directions are applied to excite the unidirectional edge modes in these systems. The spin of  dipole sources couples to the spin of evanescent waves near the edges, giving rise to the spin-momentum locked edge modes.

An anti-phase boundary is created by shifting the crystal by one-half period along the propagation direction. It can be observed in electronic systems and can be treated as a defect in the crystal that breaks the translation symmetry \cite{cohen2002structure,ahn2005electronic}. Accurate atomic manipulation is required in order to design the anti-phase boundaries in electronic systems \cite{wang2018designing,wei2014ferroelectric}. It is easier to design the anti-phase boundary in photonic system, which may help us have a deeper understanding of how the energy is distributed near the anti-phase boundary.

In this paper, we create an anti-phase boundary in a photonic crystal structure by shifting the structure along the direction of periodicity. Unidirectional propagation of the edge modes is discovered. To the authors' best knowledge, spin-momentum locked edge modes have not been found on anti-phase boundaries in quantum or optical systems. It  will not only make the existence of  the propagating edge modes along anti-phase boundaries in quantum systems possible, but also provide a new way to design chiral waveguides in photonic crystal structures. 

\begin{figure}
	\centering
	\subfloat[]{{\includegraphics[width=.2\textwidth]{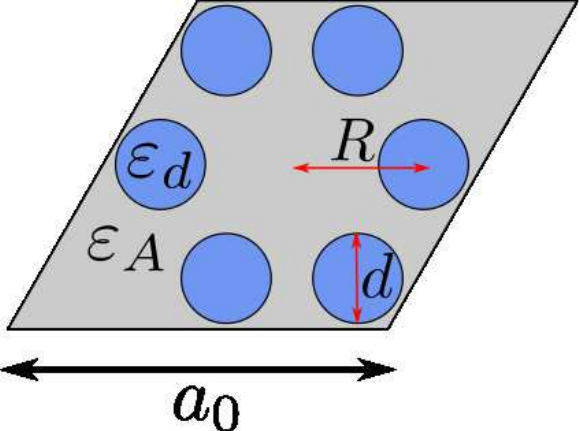}}\label{fig1a}}
	
	\subfloat[]{{\includegraphics[width=.5\textwidth]{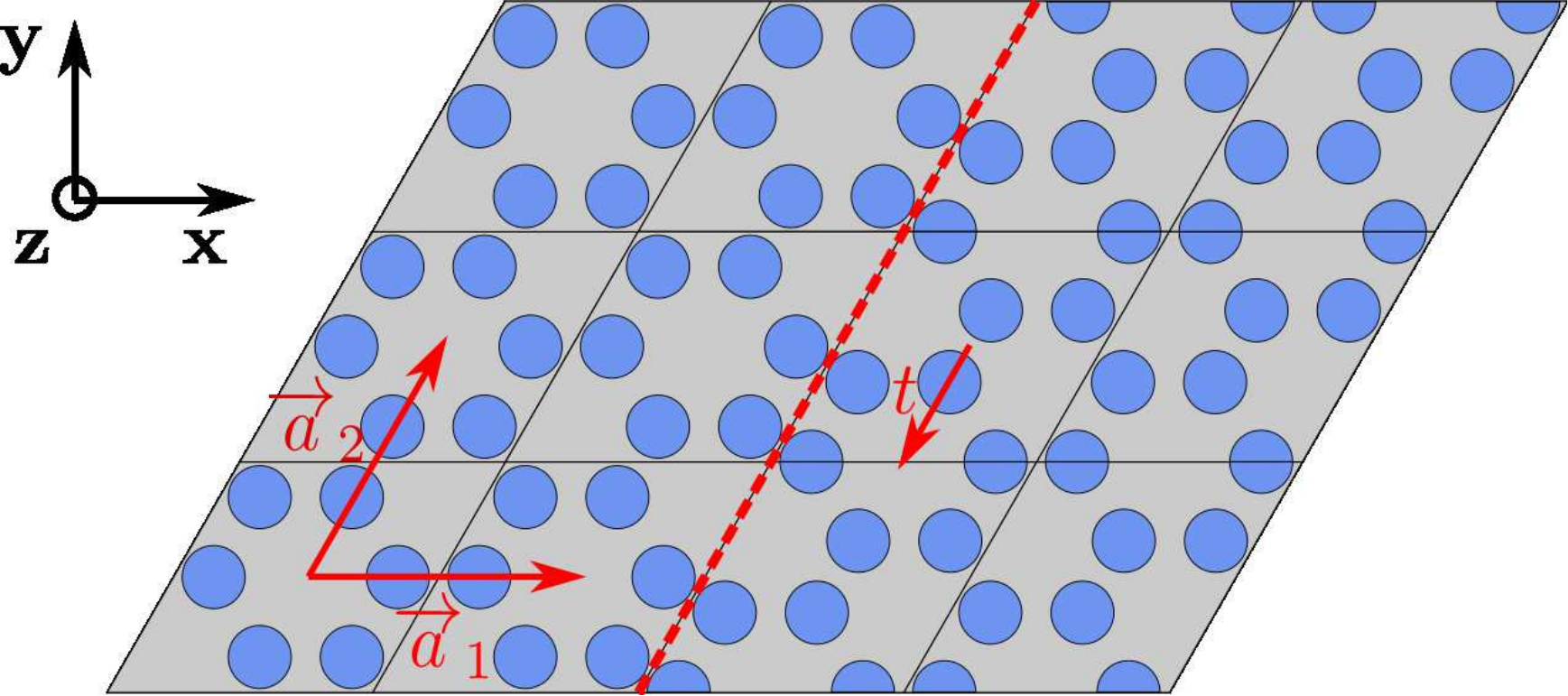}}\label{fig1b}}
	
	\caption{\label{fig1} (a) Unit cell of photonic crystal with $d$ the diameter of cylinders, $a_0$ the length of diamond edge, and $R$ the distance between the center of the diamond and the center of the cylinders. $\varepsilon_d$ and $\varepsilon_A$ are the relative permittivities of the cylinders and surrounding environment respectively. (b) Anti-phase boundary (red dashed line) formed by shifting the photonic crystal along $\protect\overrightarrow{a}_2$ by one-half period $t=-a_0/2$ where $\protect\overrightarrow{a}_1$ and $\protect\overrightarrow{a}_2$ are lattice vectors of the crystal. The angle between $\protect\overrightarrow{a}_1$ and $\protect\overrightarrow{a}_2$ is $\pi/3$. }
	
\end{figure}

\begin{figure}
	\centering
	
	\includegraphics[width=.5\textwidth]{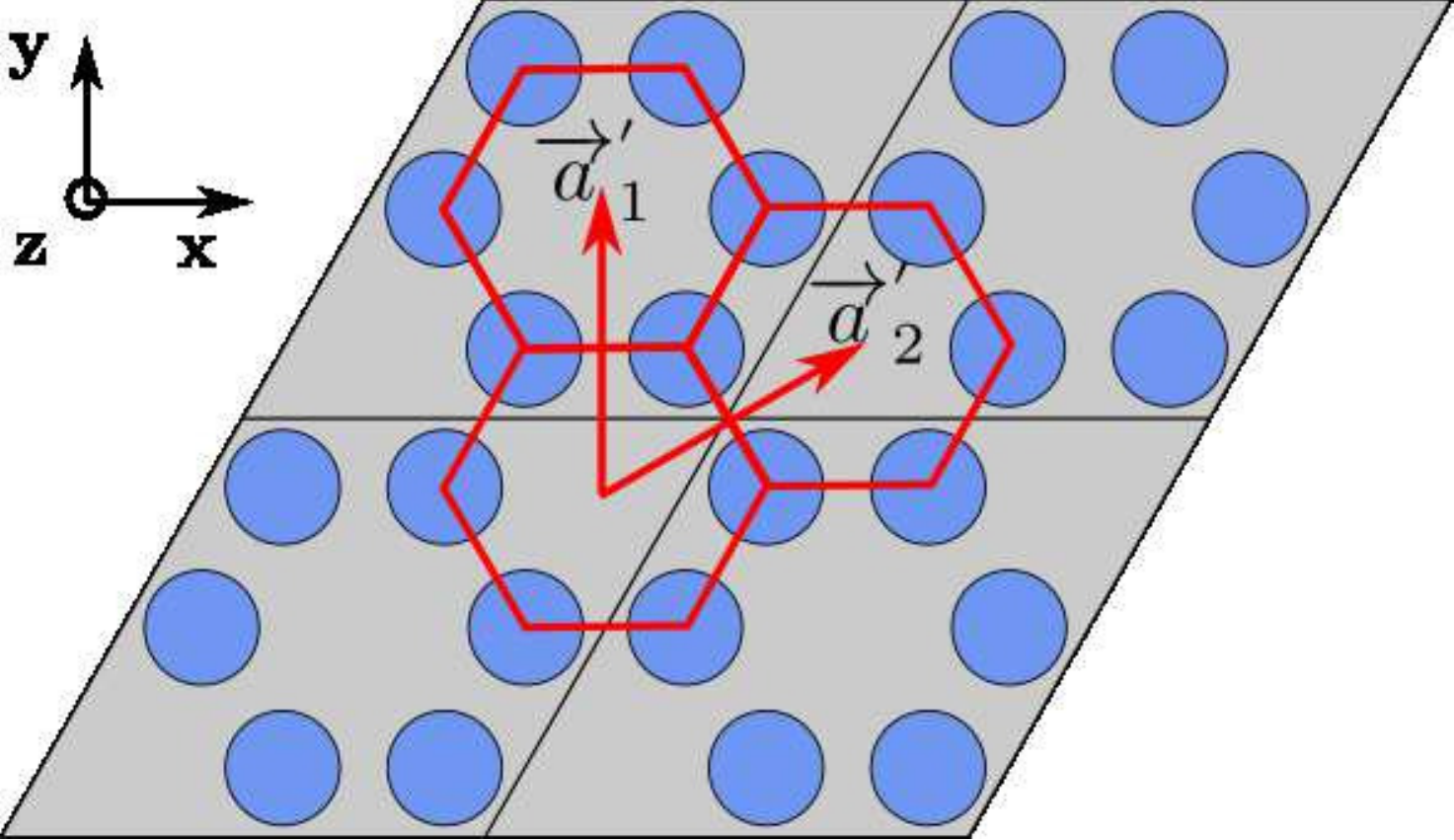}
	
	\caption{\label{fig2} The red hexagons are unit cells of the crystal for $R=a_0/3$ while $\protect\overrightarrow{a}_1^{'}$ and $\protect\overrightarrow{a}_2^{'}$ are lattice vectors. }
	
\end{figure}

\section{Spin-momentum locked modes}
\begin{figure}
	\centering
	\subfloat[]{{\includegraphics[width=.45\textwidth]{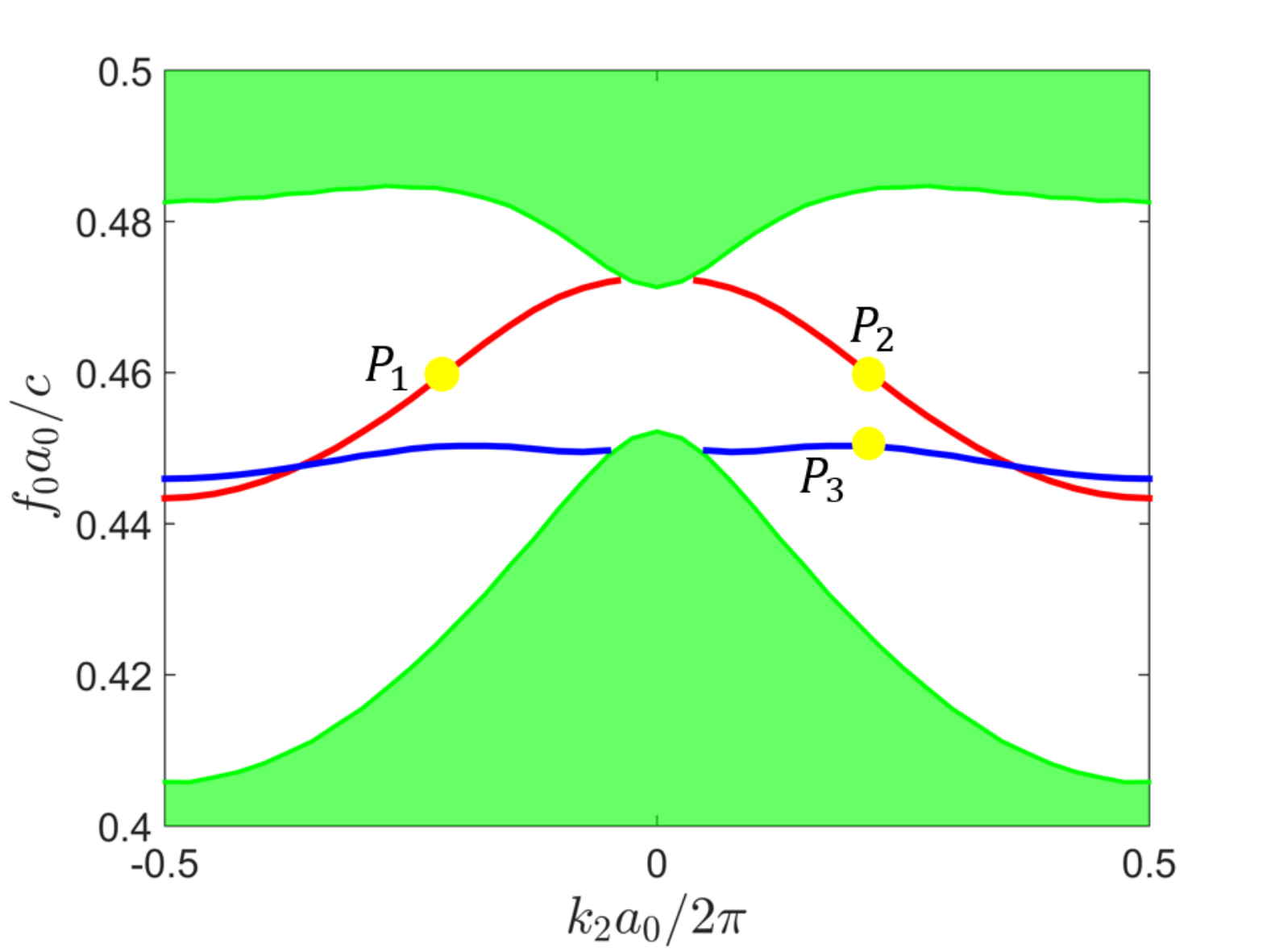}}\label{fig3a}}
	\subfloat[]{{\includegraphics[width=.55\textwidth]{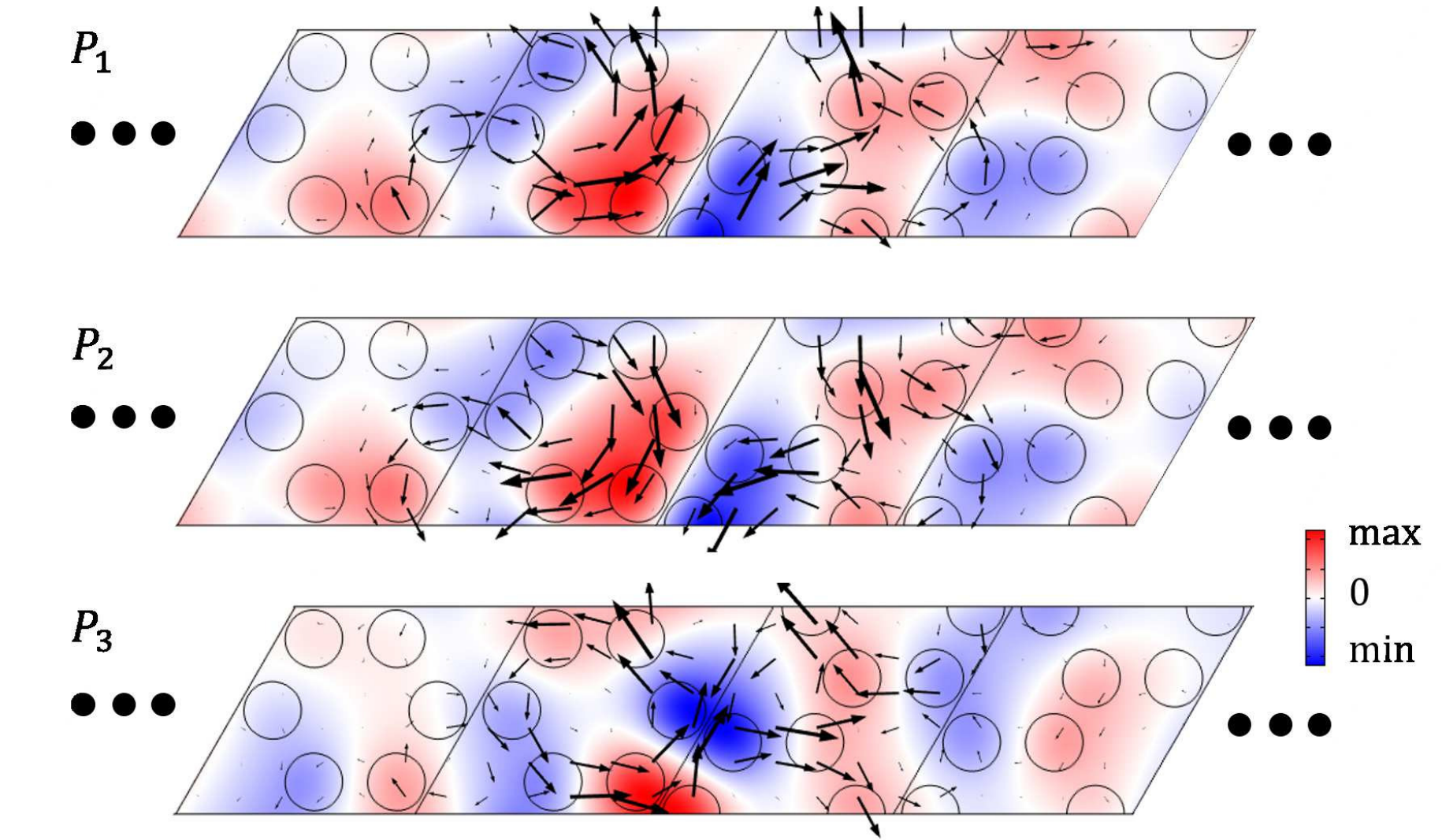}}\label{fig3b}}
	
	\subfloat[]{{\includegraphics[width=.4\textwidth]{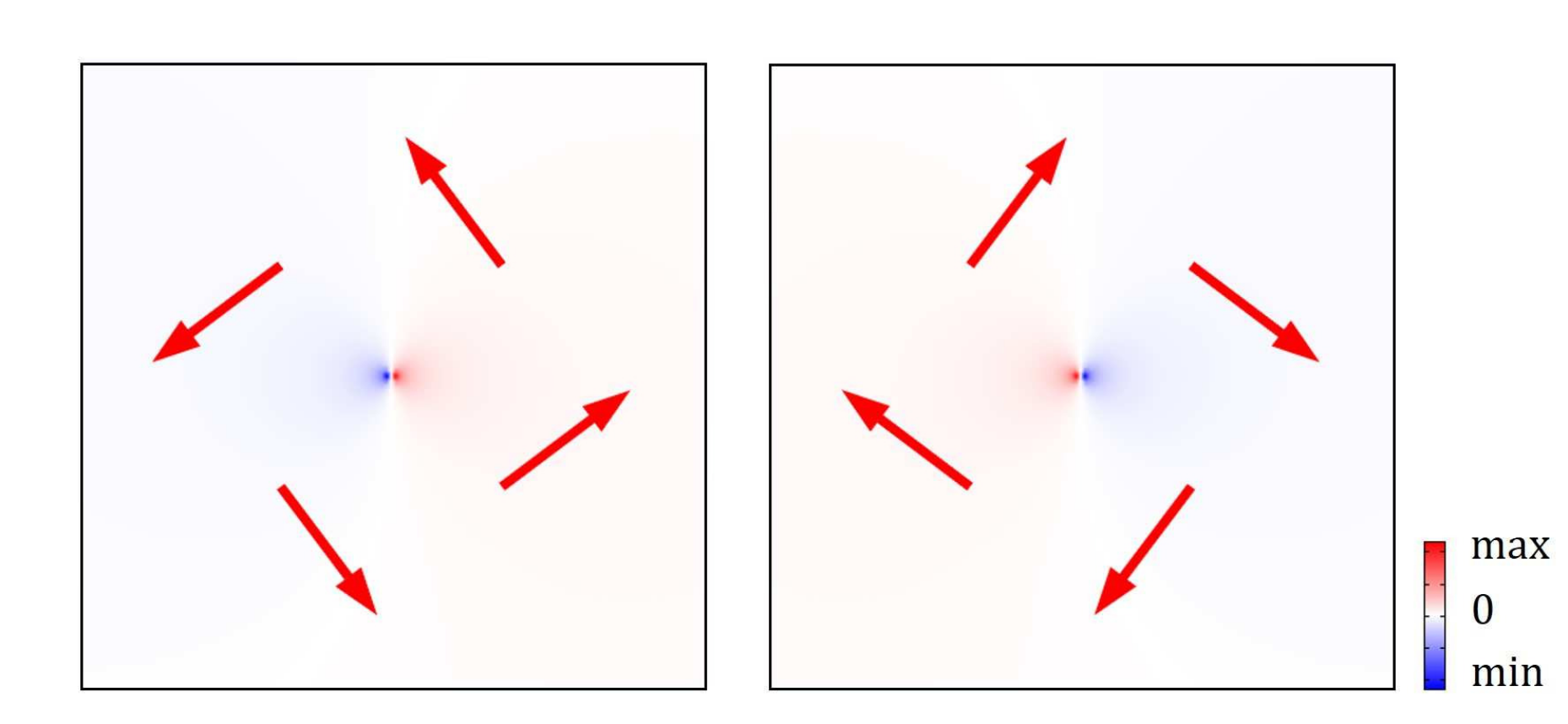}}\label{fig3c}}
	\subfloat[]{{\includegraphics[width=.65\textwidth]{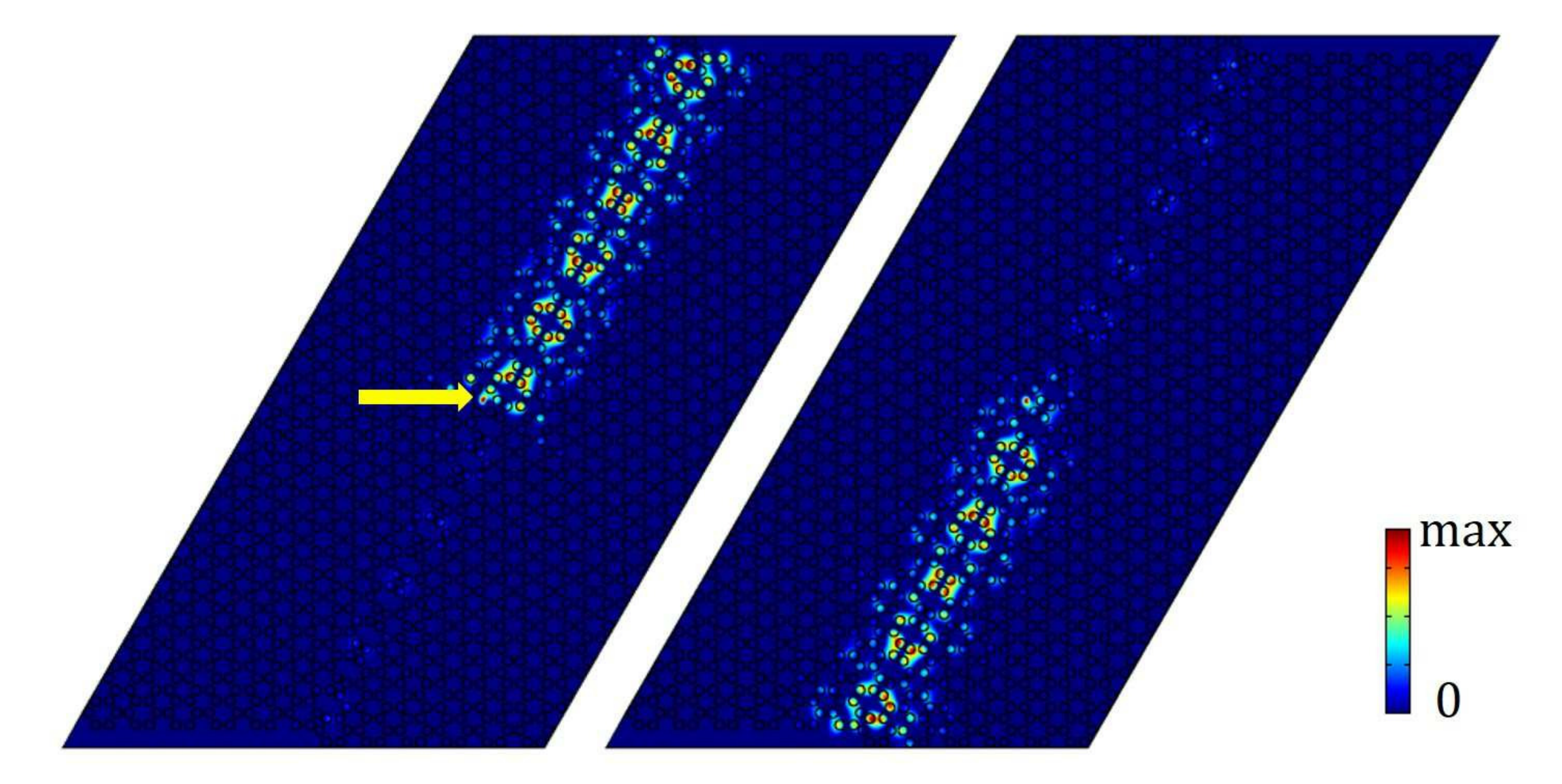}}\label{fig3d}}
	\caption{\label{fig3} (a) Dispersion relation of the super-cell which is periodic in $\protect\overrightarrow{a}_2$ direction and of 8 unit cells on each side of anti-phase boundary in $\protect\overrightarrow{a}_1$ direction. Label $k_2$ means the projection of k vector onto $\protect\overrightarrow{a}_2/|\protect\overrightarrow{a}_2|$. The green-shaded region is the projected band diagram of the bulk modes. Red and blue lines represent the odd modes and even modes respectively. The diameter of cylinder and distance between cylinder center and diamond center are   $d=0.24a_0$ and $R=0.345a_0$. The relative permittivities are $\varepsilon_d=11.7$ and $\varepsilon_A=1$. (b) Real part of $E_z$ distributions at points $P_1$, $P_2$ and $P_3$ as shown in (a). The black arrows indicate the time-averaged Poynting vectors over a period. (c) Real part distributions of $E_z$ of magnetic dipoles  $(\hat{x}-i\hat{y})/\sqrt{2}$ (left) and $(\hat{x}+i\hat{y})/\sqrt{2}$ (right) are plotted. The red arrows represent the time-averaged Poynting vectors. (d) $|E_z|$ are plotted for the driven modes excited by magnetic dipoles $(\hat{x}-i\hat{y})/\sqrt{2}$ (left) and $(\hat{x}+i\hat{y})/\sqrt{2}$ (right) respectively. The yellow arrow indicates the location of the source, which is at the center of the unit cell. The normalized frequency of the source is chosen to be $f_0a_0/c=0.46$. }
	
\end{figure}

\begin{figure}
	\centering
	\subfloat[]{{\includegraphics[width=.45\textwidth]{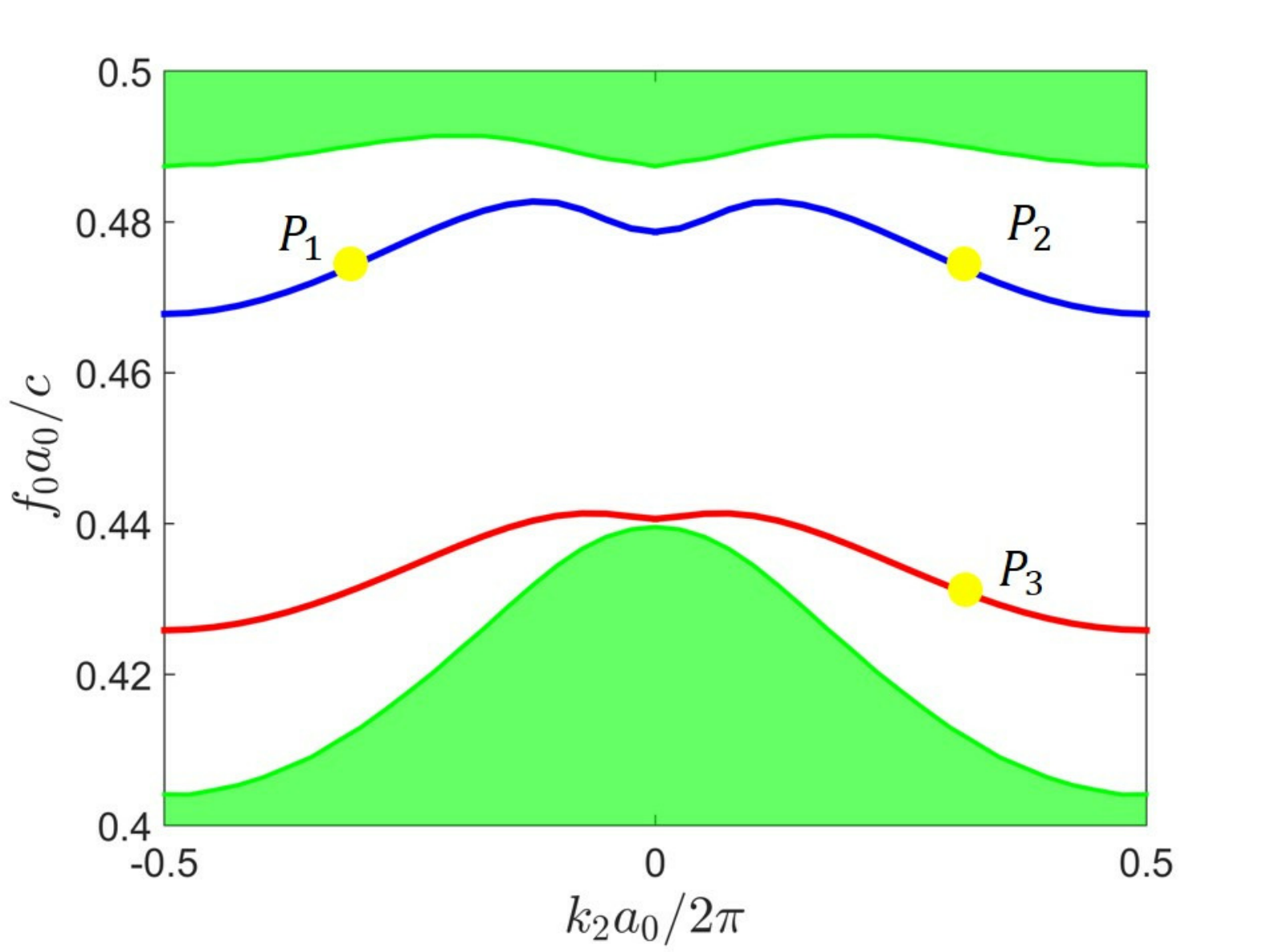}}\label{fig4a}}
	\subfloat[]{{\includegraphics[width=.55\textwidth]{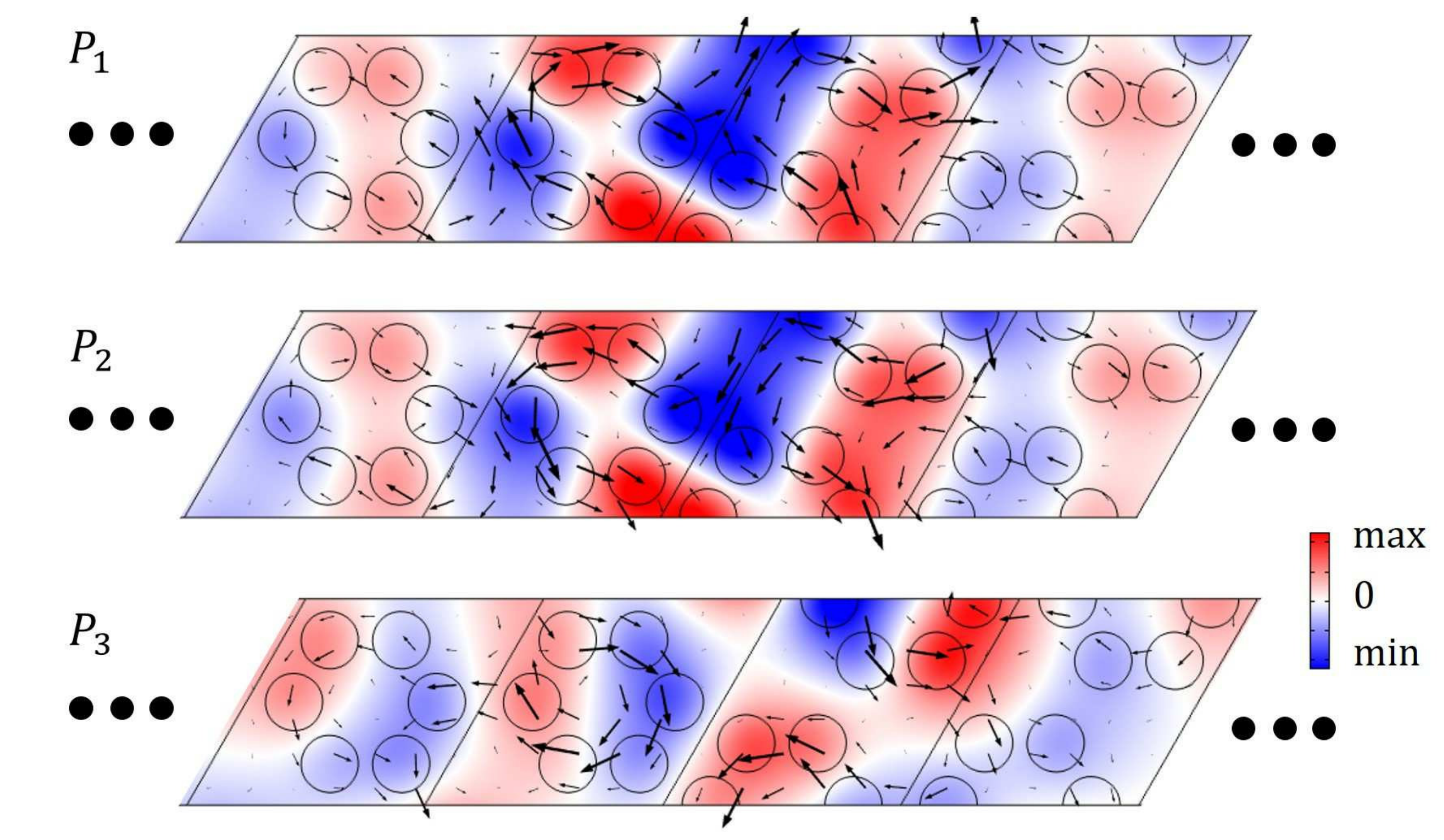}}\label{fig4b}}
	
	\subfloat[]{{\includegraphics[width=.65\textwidth]{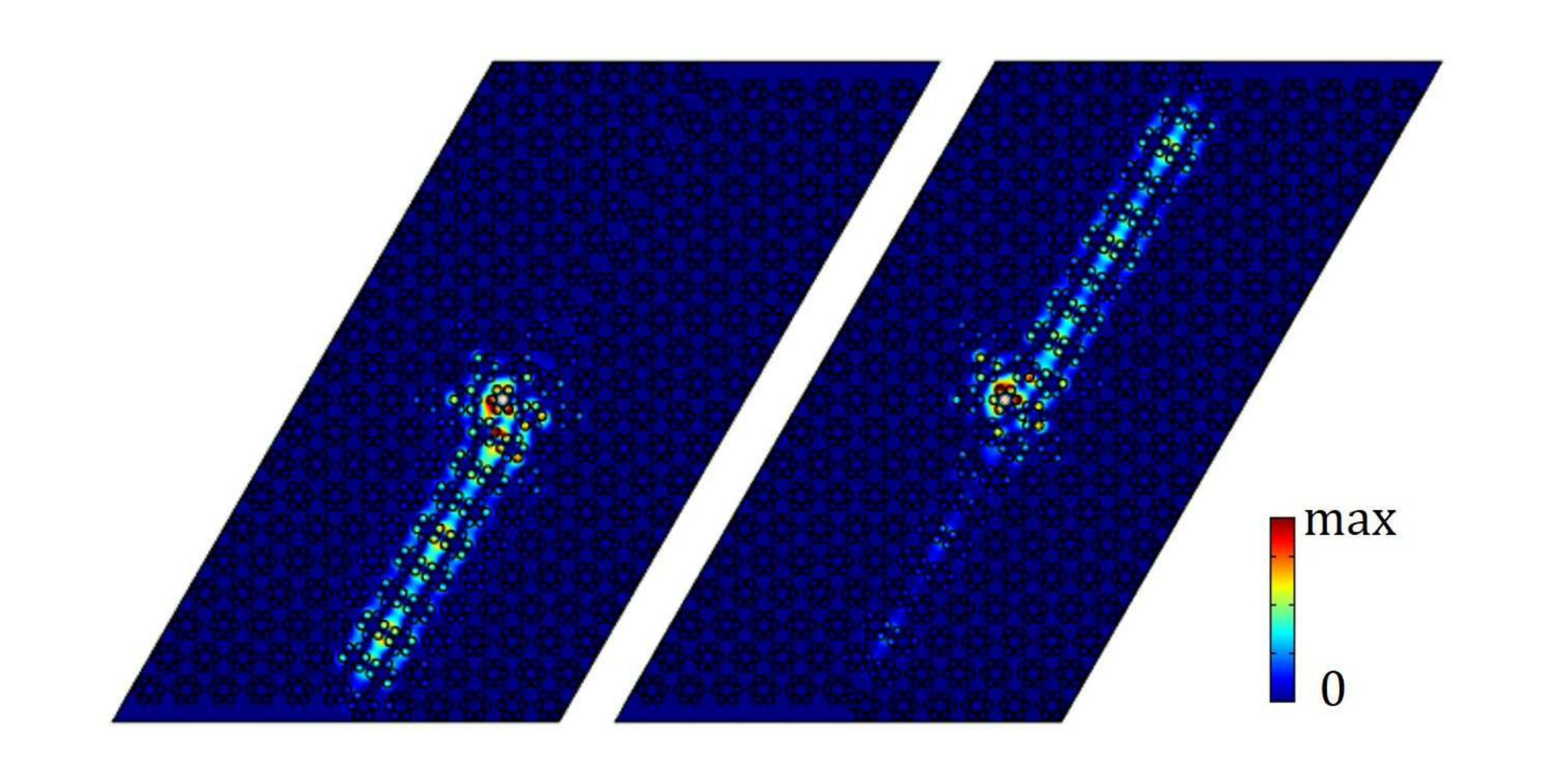}}\label{fig4c}}
	
	\caption{\label{fig4} (a) Dispersion relation of the super-cell when $R=0.3a_0$. Red and blue lines represent the odd modes and even modes respectively. (b) Real part of $E_z$ distributions at points $P_1$, $P_2$ and $P_3$ as shown in (a). (c) $|E_z|$ are plotted for the driven modes excited by magnetic dipoles $(\hat{x}-i\hat{y})/\sqrt{2}$ (left) and $(\hat{x}+i\hat{y})/\sqrt{2}$ (right) respectively. The normalized frequency of the source is chosen to be $f_0a_0/c=0.473$. }
	
\end{figure}
\begin{figure}
	\centering
	\subfloat[]{{\includegraphics[width=.5\textwidth]{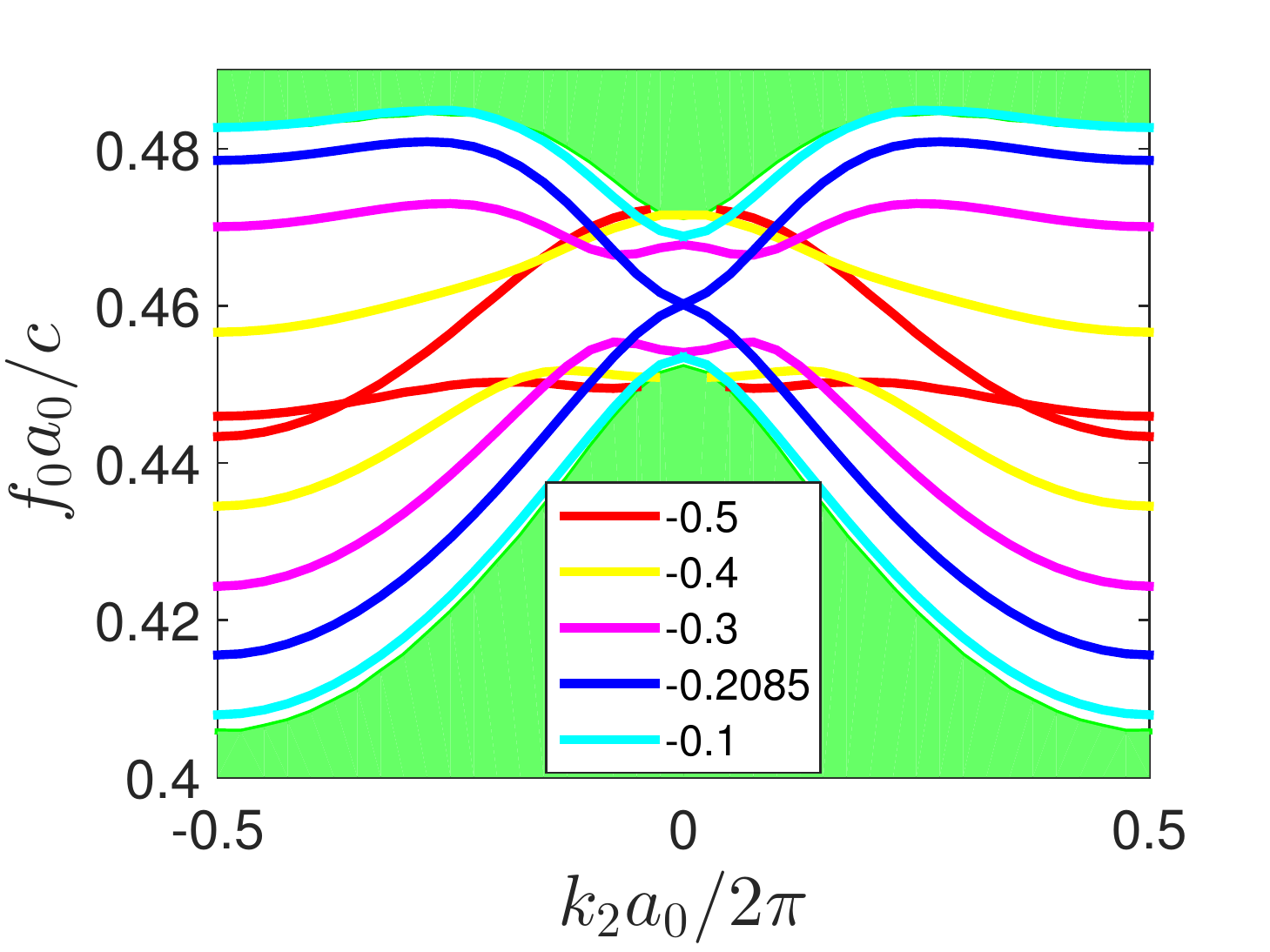}}\label{fig5a}}
	\subfloat[]{{\includegraphics[width=.475\textwidth]{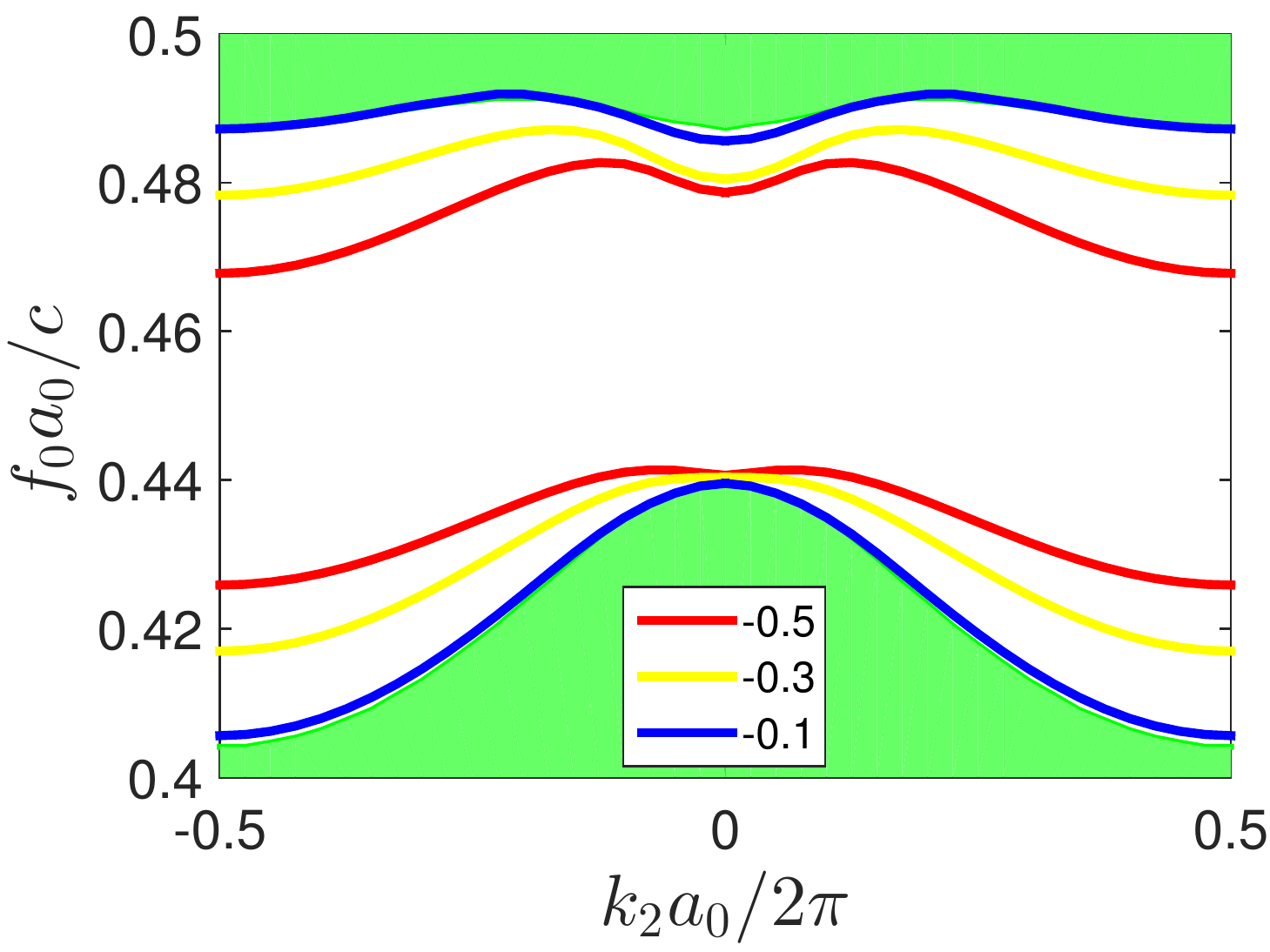}}\label{fig5b}}
	
	\subfloat[]{{\includegraphics[width=.65\textwidth]{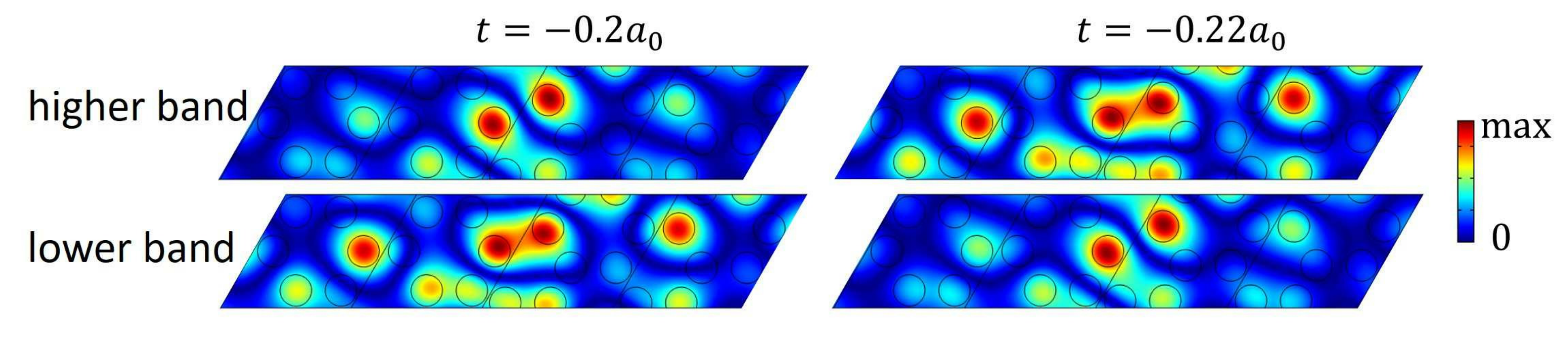}}\label{fig5c}}
	
	\subfloat[]{{\includegraphics[width=.65\textwidth]{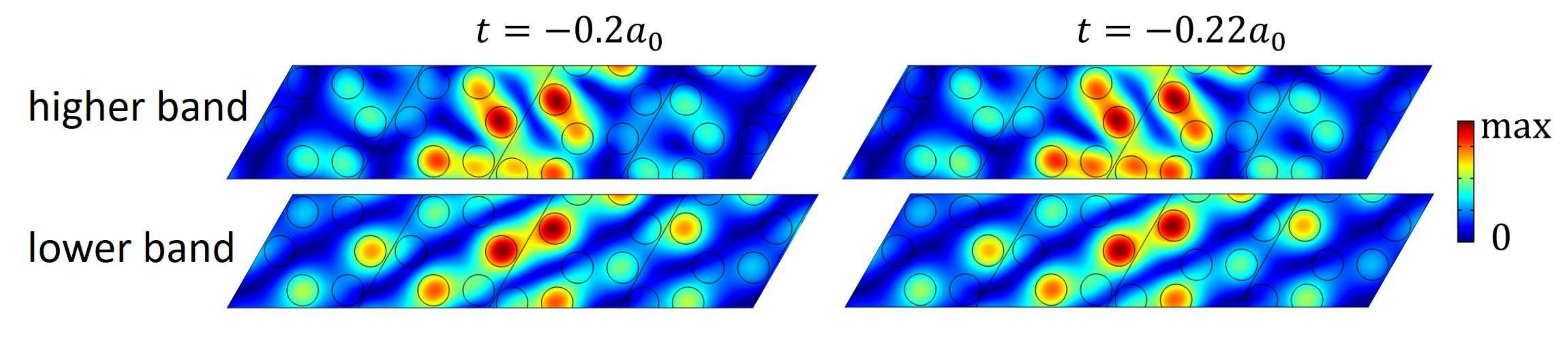}}\label{fig5d}}
	
	\caption{\label{fig5} Dispersion relations of the super-cells with (a) $R=0.345a_0$ and (b) $R=0.3a_0$ when tuning the offset $t$ in units of $a_0$. $|E_z|$ distributions are plotted for the edge modes with $R=0.345a_0$ when (c) $k_2a_0/2\pi=0$ and (d) $k_2a_0/2\pi=0.1$. }
	
\end{figure}
As shown in Fig.~\ref{fig1}, an anti-phase boundary is created by shifting the photonic crystal along the direction $\overrightarrow{a}_2$ by $-a_0/2$, which is one-half period. The geometry and material parameters are given in Fig.~\ref{fig1}. Here we only investigate the transverse magnetic (TM) modes of the electromagnetic waves, where only $E_z$, $H_x$, and $H_y$ are nonzero. According to Ref.~\cite{wu2015scheme}, tuning the distance between the center of the diamond and the center of cylinders $R$ will change the topological properties of the crystal. When $R<a_0/3$, the structure behaves as a topologically trivial material with $Z_2$ index equal to zero. Band folding occurs when $R=a_0/3$ since the lattice vectors of the unit cell change into $\overrightarrow{a}_1^{'}=-\overrightarrow{a}_1/3+2\overrightarrow{a}_2/3$ and $\overrightarrow{a}_2^{'}=\overrightarrow{a}_1/3+\overrightarrow{a}_2/3$ as shown in Fig.~\ref{fig2}. The size of the unit cell shrinks while the Brillouin zone expands. If the original Brillouin zone ($R\neq a_0/3$) is chosen, the bands on expanded Brillouin zone ($R=a_0/3$) must be folded to fit in the original one, which leads to the creation of a Dirac cone at the $\Gamma$ point. Further increasing $R$ opens the band gap at $\Gamma$ point and turns the trivial crystal into a topological insulator with nonzero $Z_2$ index. Band inversion happens at the $\Gamma$ point when $R>a_0/3$ with dipole modes in the higher band and quadrupole modes in the lower band. Unidirectional edge modes can be found at the boundary between the topological insulator ($R>a_0/3$) and trivial crystal ($R<a_0/3$). 

Here we place the topological insulator with $R>a_0/3$ on both sides of the boundary as shown in Fig.~\ref{fig1b}. However, the topological properties of the photonic crystal cannot explain the edge modes discovered on the anti-phase boundary since shifting will not change the band diagram and $Z_2$ index of the crystal. As shown in Fig.~\ref{fig3a}, the odd edge modes (anti-symmetric distributions) and the even edge modes (symmetric distributions) are caused by the mirror symmetry of the super-cell. The field distributions of the edge modes calculated by COMSOL are given in Fig.~\ref{fig3b}. The $E_z$ distributions at point $P_1$ and $P_2$ defined in Fig.~\ref{fig3a} are the same while the Poynting vectors are in opposite directions. Here we define the counter clockwise rotation of the Poynting vectors on the left side of the anti-phase boundary as spin-up and the clockwise rotation as spin-down. By comparing $P_1$ and $P_2$ we know that the edge modes with the same frequency but opposite k vectors have different spin directions. Also, we show that the edge modes with the same k vector have opposite spin directions by comparing the fields at $P_2$ and $P_3$.   

In order to excite the edge modes, a circularly polarized magnetic dipole is chosen as the source in our driven mode simulation. By observing the Poynting vectors in Fig.~\ref{fig3c}, we  conclude that magnetic dipole $(\hat{x}-i\hat{y})/\sqrt{2}$ behaves like the spin-up source while $(\hat{x}+i\hat{y})/\sqrt{2}$ like the spin-down source. The frequency of excitation is chosen to be inside the band gap of the bulk modes, which only excite the odd edge modes as we can conclude from Fig.~\ref{fig3a}. We apply the spin-up source to the shifted structure to excite the spin-up edge mode at $P_1$. Since the group velocity at $P_1$ is positive, the wave will propagate along the direction $\overrightarrow{a}_2$. The simulation result shown on the left side of Fig.~\ref{fig3d} matches this theoretical prediction. Similarly, a spin-down source will excite the edge mode propagates along $-\overrightarrow{a}_2$, which is also shown on the right side of Fig.~\ref{fig3d}.

Tuning the parameter $R$ to $R<a_0/3$ will dramatically change the properties of the edge modes. According to Ref.~\cite{wu2015scheme}, the band diagram has been closed and reopened at the $\Gamma$ point when tuning the $R$ from $R>a_0/3$ to $R<a_0/3$. The even edge mode rises while the odd mode declines. As shown in Fig.~\ref{fig4a}, the even mode is above the odd mode inside the band gap when $R=0.3a_0$, which is opposite to the result shown in Fig.~\ref{fig3a}. If we apply the spin-up source $(\hat{x}-i\hat{y})/\sqrt{2}$ with normalized frequency inside the band gap, it will excite the spin-up edge mode at $P_2$ as shown in Fig.~\ref{fig4b}. Since the group velocity at $P_2$ is negative, the wave will propagate along the $-\overrightarrow{a}_2$ direction, which is verified by the left part of Fig.~\ref{fig4c}. This indicates both topological and trivial photonic system can form anti-phase boundary and support spin-momentum locked edge modes on the boundary. The source of the same spin can excite wave with opposite propagation directions in these two photonic crystal systems.

\section{Band inversion of edge modes when tuning the offset}

By tuning the offset $t$ which is defined in Fig.~\ref{fig1b}, we can get a series of dispersion relations as shown in Fig.~\ref{fig5a} and Fig.~\ref{fig5b}. Since the mirror symmetry is broken for $t\neq-0.5a_0$, we can't define the odd mode or even mode according to the mirror plane. For the trivial unit cell, varying from the anti-phase boundary with $t=-0.5a_0$ to the two dimensional photonic crystal with $t=0$ will make the dispersion curve get closer to the projected bulk band diagram. The variation of dispersion curves for the structure consisting of topological unit cell is more complicated. As shown in Fig.~\ref{fig5a}, the two dispersion curves converge at the $\Gamma$ point and form a degenerate point at $\Gamma$ when the offset $t=-0.2085a_0$. If we continue changing $t$ from $-0.2085a_0$ to $0$, the gap between two edge modes reopens and increases until the two curves vanish into the bulk bands. 

The band inversion occurs at the $\Gamma$ point when the offset crosses over the degenerate case $t=-0.2085a_0$. As shown in Fig.~\ref{fig5c}, the $E_z$ distributions in the higher band of the case $t=-0.2a_0$ are the same as the lower band when $t=-0.22a_0$. When $k_2$ is sufficiently far away from the $\Gamma$ point, the field distributions look similar in the higher band or lower band for different offsets. We can conclude that only the edge modes that are close to $\Gamma$ point will be inverted when $-0.5a_0<t<0.2085a_0$, which is similar to the band inversion of the bulk modes in Ref.~\cite{wu2015scheme}. 

\section{Edge modes in gradual shift structure}
\begin{figure}
	\centering
	\subfloat[]{{\includegraphics[width=.9\textwidth]{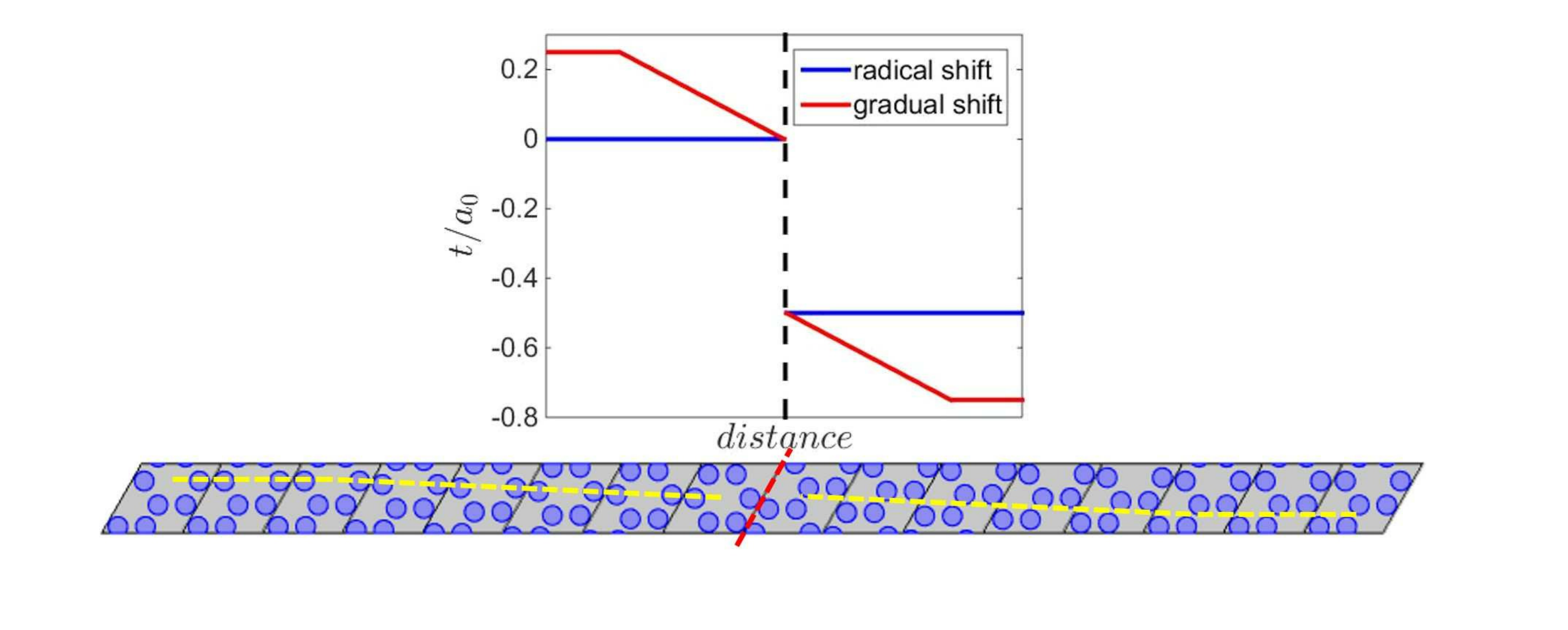}}\label{fig6a}}
	
	\subfloat[]{{\includegraphics[width=.45\textwidth]{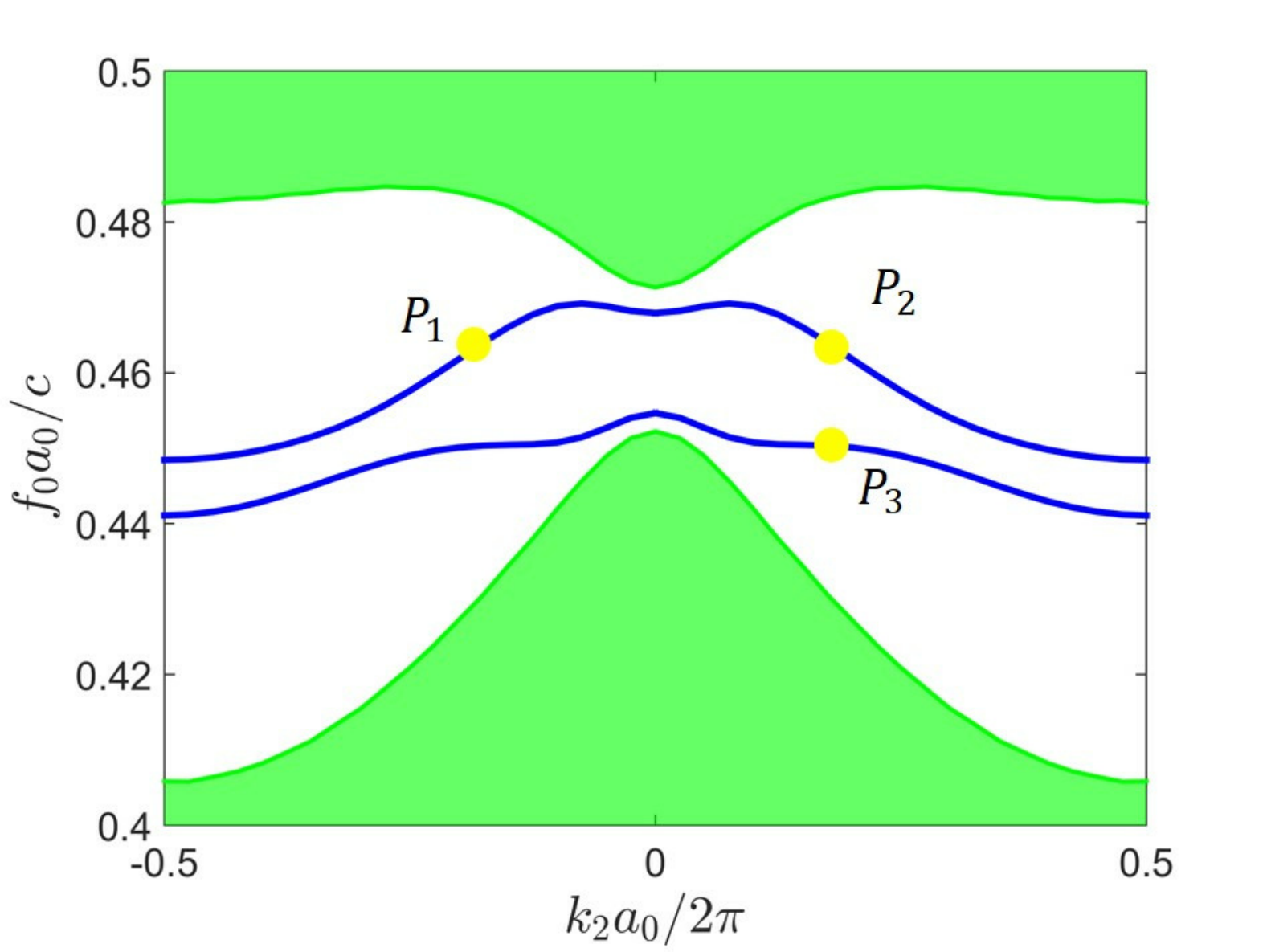}}\label{fig6b}}
	\subfloat[]{{\includegraphics[width=.55\textwidth]{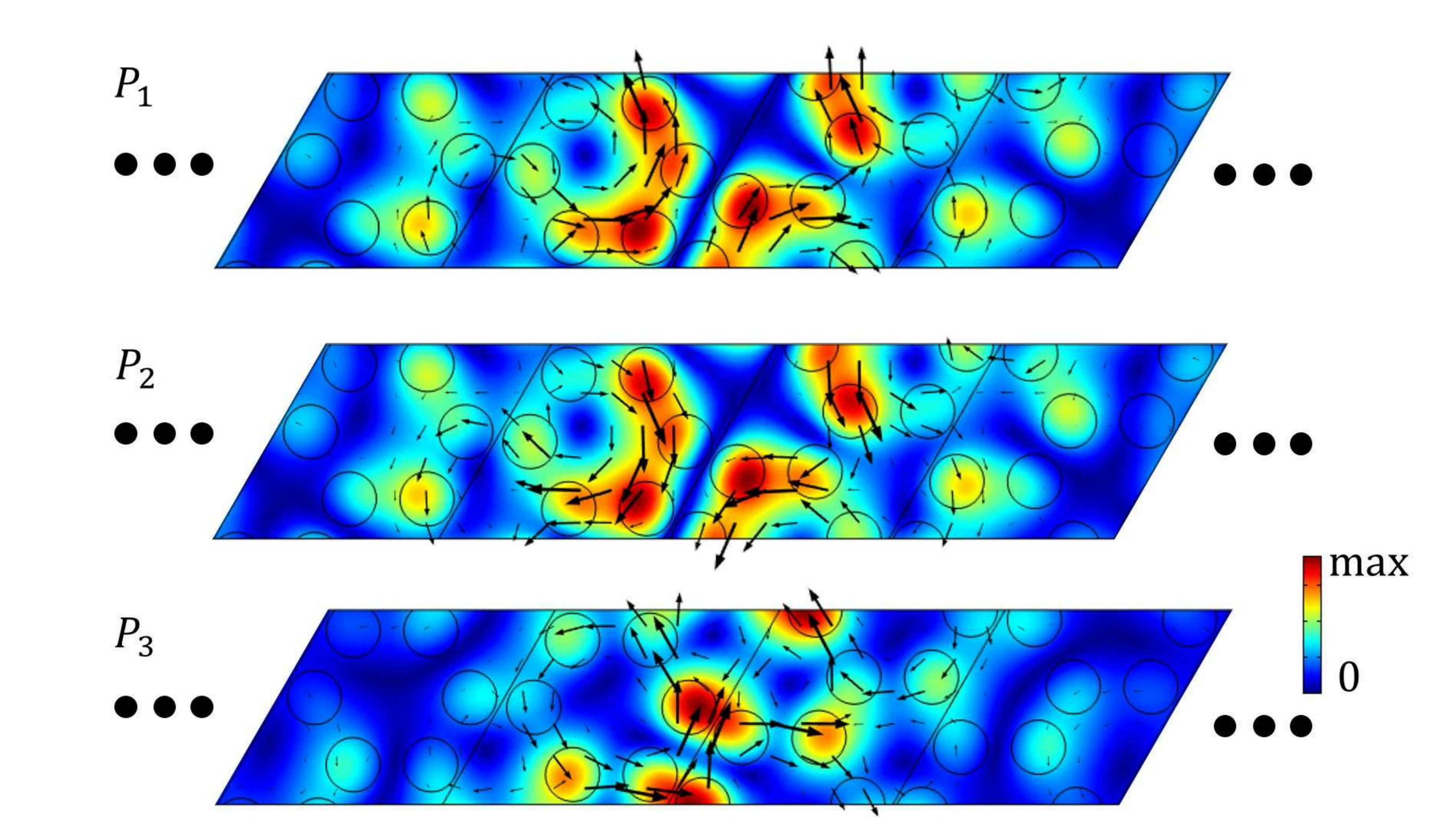}}\label{fig6c}}
	
	\subfloat[]{{\includegraphics[width=.65\textwidth]{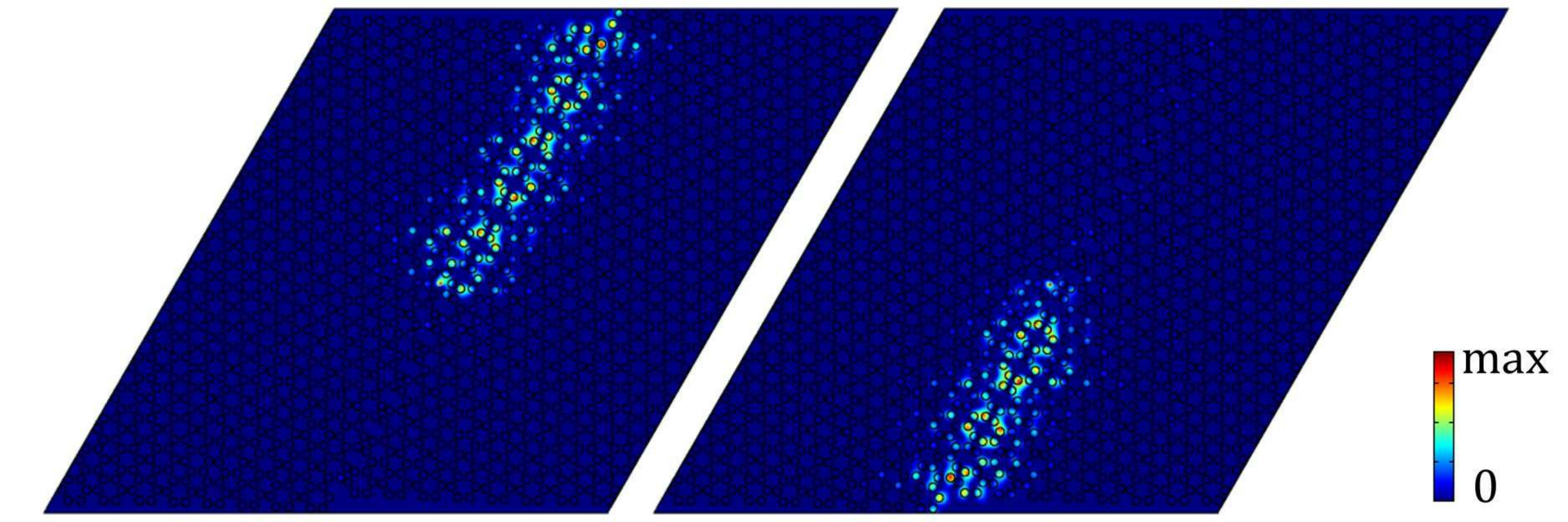}}\label{fig6d}}
	
	\caption{\label{fig6} (a) Comparison between radical and gradual shift super-cell. A shift of $t=0.05a_0$ is set between adjacent unit cells on the two sides of the boundary marked by red dashed line. The far left with $t=0.25a_0$ and the far right with $t=-0.75a_0$ have the same  pattern. (b) Dispersion relation of the gradual shift super-cell when $R=0.345a_0$. (c)  $|E_z|$ distributions at points $P_1$, $P_2$ and $P_3$ as shown in (b). (d) $|E_z|$ are plotted for the driven modes excited by magnetic dipoles $(\hat{x}-i\hat{y})/\sqrt{2}$ (left) and $(\hat{x}+i\hat{y})/\sqrt{2}$ (right) respectively. The normalized frequency of the source is chosen to be $f_0a_0/c=0.464$. The source is located at the center of the unit cell with $t=0$. }
	
\end{figure}

We can also create an anti-phase boundary by gradually shifting the unit cells on the two sides of the boundary as shown in Fig.~\ref{fig6a}.  Here the unit cell with $R=0.345a_0$ is studied. We can conclude from the dispersion relations shown in Fig.~\ref{fig5a} that the edge modes which decay rapidly into the bulk can only be found when the offset between the adjacent unit cells is large enough. For the offset with $|t|<0.1a_0$, the dispersion curves are so close to the bulk band diagram that their energy is not well confined to the boundary. Hence the offset of $t=0.05a_0$ is chosen between the adjacent unit cells on the two sides of the anti-phase boundary to prevent the appearance of redundant edge modes. The unit cells will look the same if they are far enough from the boundary, which is different from the radical shift structure where the offset difference always exists on the two sides. In this structure, there is no long-range offset between the two sides, only a local shift in the unit cells near the boundary. The dispersion relation and field distribution are shown in Fig.~\ref{fig6b} and Fig.~\ref{fig6c} respectively, which is similar to the radical shift case as shown in Fig.~\ref{fig3a} and Fig.~\ref{fig3b}. The unidirectional propagation of the edge modes can also be found when we excite with sources of different spin directions as shown in Fig.~\ref{fig6d}.

\section{Optimization of the source}
\begin{figure}
	\centering
	\subfloat[]{{\includegraphics[width=.5\textwidth]{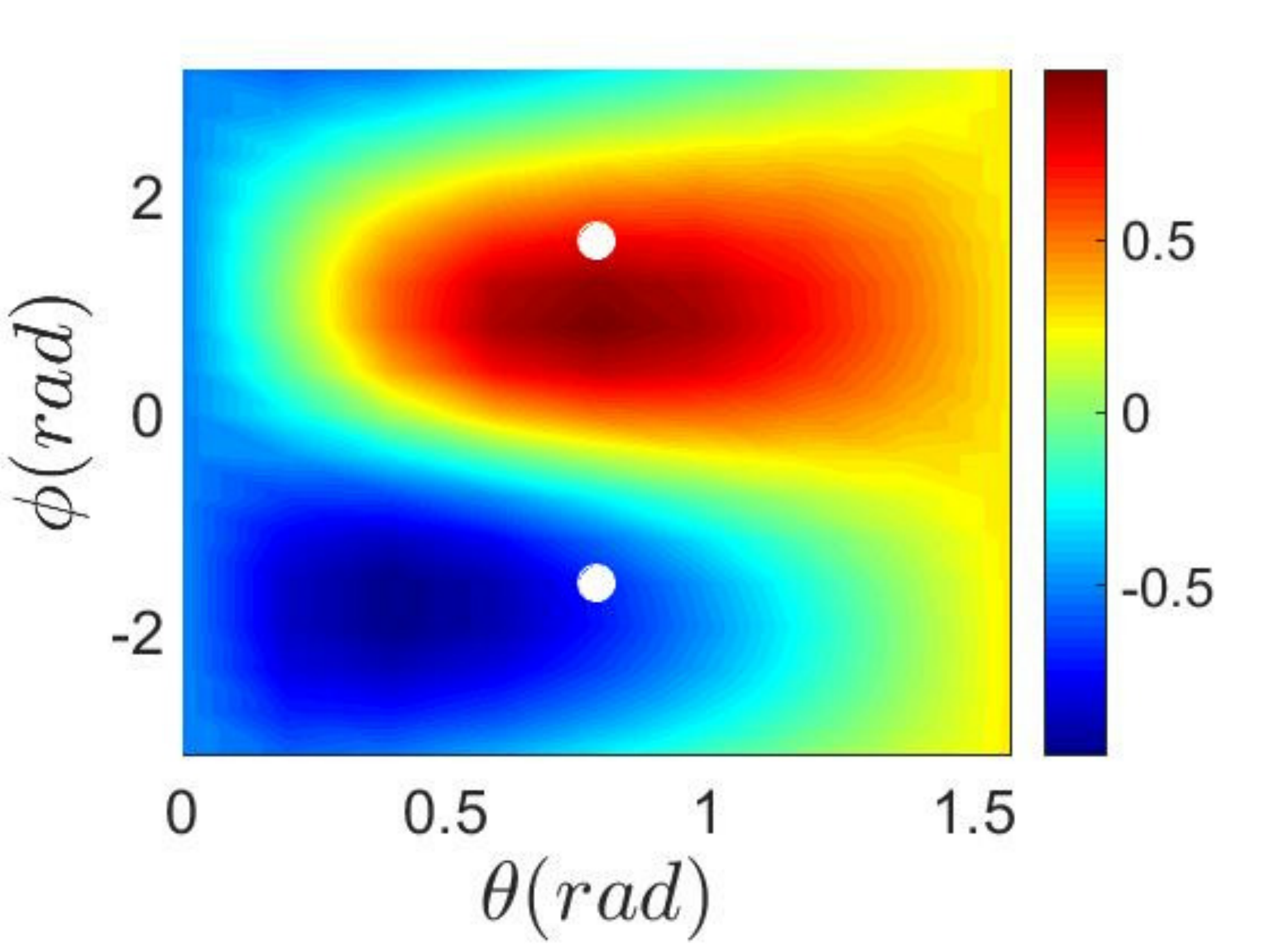}}\label{fig7a}}
	\subfloat[]{{\includegraphics[width=.5\textwidth]{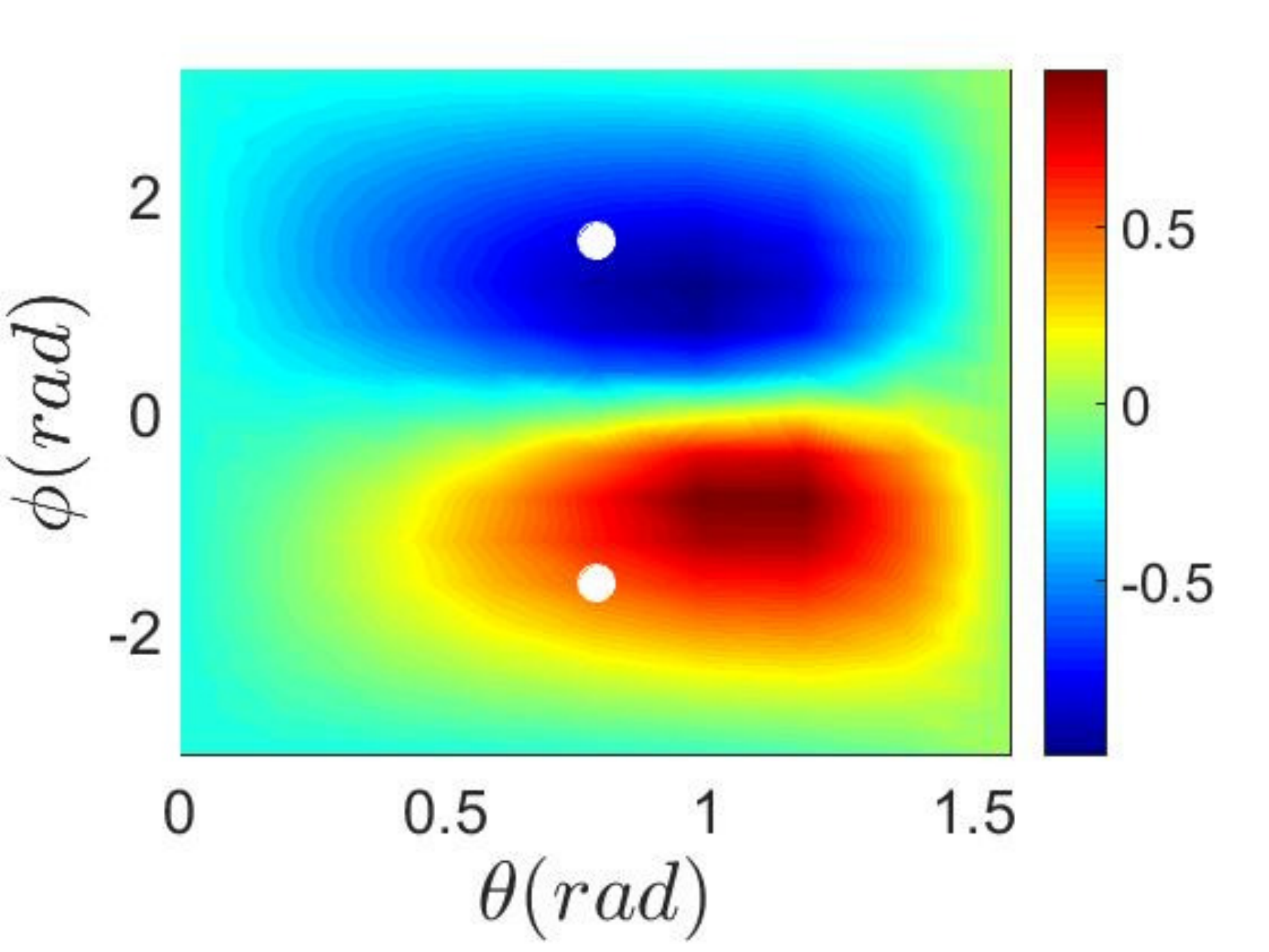}}\label{fig7b}}

	\caption{\label{fig7} Directionality $D$ defined in Eq.~\ref{eq:direction} is plotted as a function of $\theta$ and $\phi$ for (a) $R=0.345a_0, f_0a_0/c=0.46$ and (b) $R=0.30a_0, f_0a_0/c=0.473$. The white dots indicate the locations of the sources $(\hat{x}-i\hat{y})/\sqrt{2}$ (upper) and $(\hat{x}+i\hat{y})/\sqrt{2}$ (lower). }
	
\end{figure}
By optimizing the combination of two orthogonal magnetic dipoles, we can achieve edge modes with better directionality. The magnetic dipole can be defined as:
\begin{equation}
\overrightarrow{m}=\cos\theta\hat{x}+\sin\theta\exp(-i\phi)\hat{y}
\label{eq:source}
\end{equation}
where $0<\theta<\pi/2$ and $-\pi<\phi<\pi$.
The spin-up ($(\hat{x}-i\hat{y})/\sqrt{2}$) and spin-down ($(\hat{x}+i\hat{y})/\sqrt{2}$) source mentioned above are the particular cases when $\theta$ and $\phi$ in Eq.~\ref{eq:source} are set to $\pi/4, \pi/2$ and $\pi/4, -\pi/2$ respectively. 

According to Ref.~\cite{petersen2014chiral}, we can also define the directionality of the edge mode by
\begin{equation}
D=\frac{c_{+}-c_{-}}{c_{+}+c_{-}}
\label{eq:direction}
\end{equation}
where $c_{+}$($c_{-}$) is the line integration of the Poynting vector measured on the top(bottom) of the structure as shown in Fig.~\ref{fig4c}. If $|D|$ is close to 1, we can conclude that the system has good directionality while no directionality can be observed when $D=0$. As shown in Fig.~\ref{fig7}, the signs of $D$ at the locations of  spin-up and spin-down source are opposite for $R>a_0/3$ and $R<a_0/3$, which verifies the conclusion that wave propagates in opposite directions for the same source when we tune the $R$ of the system.

\section{Conclusion}
Spin-momentum locked edge modes are discovered on the anti-phase boundaries which are formed by shifting two halves of a photonic crystal along the direction of periodicity. By applying magnetic dipole sources with different spin directions, we can excite the edge modes propagating in opposite directions. The inversion of the edge modes is revealed when we adjust the distance between the center of the unit cell and the cylinders, which leads to opposite propagation directions with the same source. Also, tuning the offset of the unit cells on two sides can cause band inversion of the edge modes for the topologically non-trivial photonic crystal system. Optimization of the source gives the edge modes better directionality and helps us to further understand the system, making it more practical for the unidirectional wave propagation applications.

\section*{Acknowledgments}
This work was supported in part by Air Force Office of Scientific Research Grant No. FA9550-16-1-0093 and in part by the China Scholarship Council (No. 201706230113). The authors acknowledge
discussions with D. Bisharat.

\bibliographystyle{unsrt}  





\bibliography{template}

\end{document}